\documentclass[aip,reprint]{revtex4-1}
\usepackage{bm}
\usepackage{graphicx}
\usepackage{makeidx}
\usepackage{amsmath,amssymb}

\newcommand{\pa}[0]{
\setlength\unitlength{0.005truecm}
\begin{picture}(120,100)(-10,31)
\linethickness{5\unitlength}
\put(0,0){\framebox( 90, 90){}}
\linethickness{25\unitlength}
\put(0,10){\line(1,0){90}}
\put(0,80){\line(1,0){90}}
\end{picture}
}
\newcommand{\pb}[0]{
\setlength\unitlength{0.005truecm}
\begin{picture}(120,100)(-10,31)
\linethickness{5\unitlength}
\put(0,0){\framebox( 90, 90){}}
\linethickness{25\unitlength}
\put(10,0){\line(0,1){90}}
\put(80,0){\line(0,1){90}}
\end{picture}
}
\newcommand{\pc}[0]{
\setlength\unitlength{0.005truecm}
\begin{picture}(120,100)(-10,31)
\linethickness{5\unitlength}
\put(0,0){\framebox( 90, 90){}}
\linethickness{25\unitlength}
\put(0,80){\line(1,0){90}}
\end{picture}
}
\newcommand{\pd}[0]{
\setlength\unitlength{0.005truecm}
\begin{picture}(120,100)(-10,31)
\linethickness{5\unitlength}
\put(0,0){\framebox( 90, 90){}}
\linethickness{25\unitlength}
\put(80,0){\line(0,1){90}}
\end{picture}
}
\newcommand{\pe}[0]{
\setlength\unitlength{0.005truecm}
\begin{picture}(120,100)(-10,31)
\linethickness{5\unitlength}
\put(0,0){\framebox( 90, 90){}}
\linethickness{25\unitlength}
\put(0,10){\line(1,0){90}}
\end{picture}
}
\newcommand{\pf}[0]{
\setlength\unitlength{0.005truecm}
\begin{picture}(120,100)(-10,31)
\linethickness{5\unitlength}
\put(0,0){\framebox( 90, 90){}}
\linethickness{25\unitlength}
\put(10,0){\line(0,1){90}}
\end{picture}
}
\newcommand{\pg}[0]{
\setlength\unitlength{0.005truecm}
\begin{picture}(120,100)(-10,31)
\linethickness{5\unitlength}
\put(0,0){\framebox( 90, 90){}}
\linethickness{25\unitlength}
\end{picture}
}
\draft 

\begin{document}


\title{Novel constructive method for the quantum dimer model in spin-1/2 Heisenberg antiferromagnets with frustration on a diamond-like-decorated square lattice} 



\author{Yuhei Hirose}
\email{y.hirose@rs.tus.ac.jp}
\author{Akihide Oguchi}
\author{Masafumi Tamura}
\author{Yoshiyuki Fukumoto}
\email{yfuku@rs.tus.ac.jp}
\affiliation{Tokyo University of Science, Noda, Chiba 278-8510, Japan}


\date{\today}

\begin{abstract}
We study spin-1/2 Heisenberg antiferromagnets on a diamond-like-decorated square lattice. 
The diamond-like-decorated square lattice is a lattice in which the bonds in a square lattice are replaced with diamond units. 
The diamond unit has two types of antiferromagnetic exchange interactions, and the ratio $\lambda$ of the diagonal bond strength to that of the other four edges controls the frustration strength. 
For $0.974<\lambda<2$, the present system has a nontrivial macroscopic degeneracy, which is called the macroscopically degenerated tetramer$-$dimer (MDTD) states. 
The MDTD states are identical to the Hilbert space of the Rokhsar--Kivelson (RK) quantum dimer model (QDM).  
By introducing further neighbor couplings in the MDTD states, we calculate the  second-order effective Hamiltonian, which is exactly the same as the square-lattice QDM with a finite hopping amplitude $t$ and dimer-dimer interaction $v$.
Furthermore, we calculate $v/|t|$ as a function of the ratio $\lambda$ in the Heisenberg model and examine which phases of the square-lattice QDM appear in our obtained states. 
Our obtained QDM has a region where $\lambda$ exhibits a finite hopping amplitude ($|t|>0$) and repulsive interaction between dimers ($v>0$). This suggests the possibility of realizing the resonating valence bond (RVB) state because the RVB state is obtained at $v=|t|$, which is known as the RK point. 
\end{abstract}

\pacs{}

\maketitle 


\section{Introduction}
\label{sect:1}

Exploration of the resonating valence bond (RVB) state is one of the central issues in condensed matter physics.\cite{Anderson1973}
In 1988, Rokhsar and Kivelson, motivated by the discovery of a high-temperature cuprate superconductor, proposed the quantum dimer model (QDM)\cite{Rokhsar1988} as a phenomenological Hamiltonian within RVB theory.
The QDM is expressed as 
\begin{align}
H_{\rm{QDM}}=&-t\sum \Bigl( \Bigr| \pa{} \Bigl\rangle \Bigl\langle\pb{}\Bigr|+\Bigr| \pb{} \Bigl\rangle \Bigl\langle\pa{}\Bigr| \Bigr)\notag \\
		      &+v\sum \Bigl( \Bigr| \pa{} \Bigl\rangle \Bigl\langle\pa{}\Bigr|+\Bigr| \pb{} \Bigl\rangle \Bigl\langle\pb{}\Bigr| \Bigr),
\label{eq:1}
\end{align}
where $t$ and $v$ represent the pair-hopping amplitude and dimer-dimer interaction, respectively. 
Because the RVB state emerges at $v=|t|$, known as the RK point, much effort has been devoted to constructing the QDM from quantum spin systems for the purpose of discovering the RVB states.\cite{book} 
However, it has not been made clear whether the QDM can be realized from realistic quantum spin systems. 
Thus, discovery of a quadratic and simple spin Hamiltonian that yields the QDM would be a significant step facilitating exploration of the RVB state. 
Therefore, we show that the QDM can be realized as a low-energy effective Hamiltonian of a spin-1/2 Heisenberg antiferromagnet on a diamond-like-decorated square lattice, as shown in Fig.~1. 
In this way, we present a constructive method to realize the QDM from the spin system.\cite{Hirose1,Hirose2}

A diamond-like-decorated square lattice is a lattice in which the square lattice bonds are replaced with diamond units,\cite{Strecka} as shown in Fig.~1. 
If we define the interaction strength of the four sides of a diamond unit as $J$ and that of the diagonal bond as $J'=\lambda J$, 
the ratio $\lambda$ determines the ground-state properties.\cite{Hirose3}
As shown in Fig.~1, we denote the four $S=1/2$ operators in a diamond unit as $\bm{s}_i$, $\bm{s}_j$, $\bm{s}_{k,a}$, and $\bm{s}_{k,b}$. 
Thus, the Hamiltonian can be written as 
\begin{equation} 
H=J\sum_{\langle i,j \rangle}\left\{(\bm{s}_i+\bm{s}_j)\cdot(\bm{s}_{k,a}+\bm{s}_{k,b})+\lambda \left(\bm{s}_{k,a}\cdot\bm{s}_{k,b}+\frac{3}{4}\right)\right\}
  \label{eq:2},
\end{equation}
where $\langle i,j \rangle$ represents a nearest-neighbor pair of the square lattice. 
Here, we call $\bm{s}_i$ and $\bm{s}_j$ the edge spins (closed circles in Fig.~1) and the pair ($\bm{s}_{k,a}, \bm{s}_{k,b}$) a bond spin-pair (open circles). 
The ground state of an isolated diamond unit for $\lambda<2$ becomes a nonmagnetic tetramer-singlet state, which is described by
\begin{align}
   |\phi^g\rangle_{i,j,k}=&\frac{1}{\sqrt{3}}
   \Bigl(
   |\!\uparrow\uparrow\rangle_{i,j}|t^-\rangle_k+|\!\downarrow\downarrow\rangle_{i,j}|t^+\rangle_k
   \notag\\
   & -\frac{|\!\uparrow\downarrow\rangle_{i,j}+|\!\downarrow\uparrow\rangle_{i,j}}{\sqrt{2}}|t^0\rangle_k
   \Bigr),
\label{eq:3} 
\end{align}
where $\{|t^+\rangle, |t^0\rangle, |t^-\rangle\}$ represents the triplet states of a bond spin-pair. 
The tetramer-singlet state is equivalent to the plaquette RVB state.
For a diamond-like-decorated square lattice with $0.974<\lambda<2$, the ground-state manifold consists of macroscopically degenerated tetramer-dimer (MDTD) states,\cite{Hirose3,Morita} as shown in Fig.~2(a), where the phase boundary $\lambda=0.974$ is calculated by the modified spin-wave method.\cite{Hirose3} 
If we regard a tetramer singlet as a ``dimer'' in the QDM, then MDTD states are equivalent to square-lattice dimer-covering states, as shown in Fig.~2(b).
Furthermore, we derive a square-lattice QDM as a second-order effective Hamiltonian by introducing the further neighbor couplings, and we calculate $v/|t|$ as a function of the ratio $\lambda$ in the Heisenberg model. 
Our obtained QDM has a region of $\lambda$ with a finite hopping amplitude ($|t|>0$) and a repulsive interaction between dimers ($v>0$).
This suggests the possibility of realizing the RVB state, because this state is obtained at $v=|t|$.

This paper is organized as follows. We define the second-order effective Hamiltonian and the square-lattice QDM in Sect.~2. 
In Sect.~3, we show the calculated dependence of $v/|t|$ on $\lambda$ by introducing the further neighbor couplings. 
We summarize the results obtained in this study in Sect.~4. 
\begin{figure}[t] 
\begin{center}
\includegraphics[width=.75\linewidth]{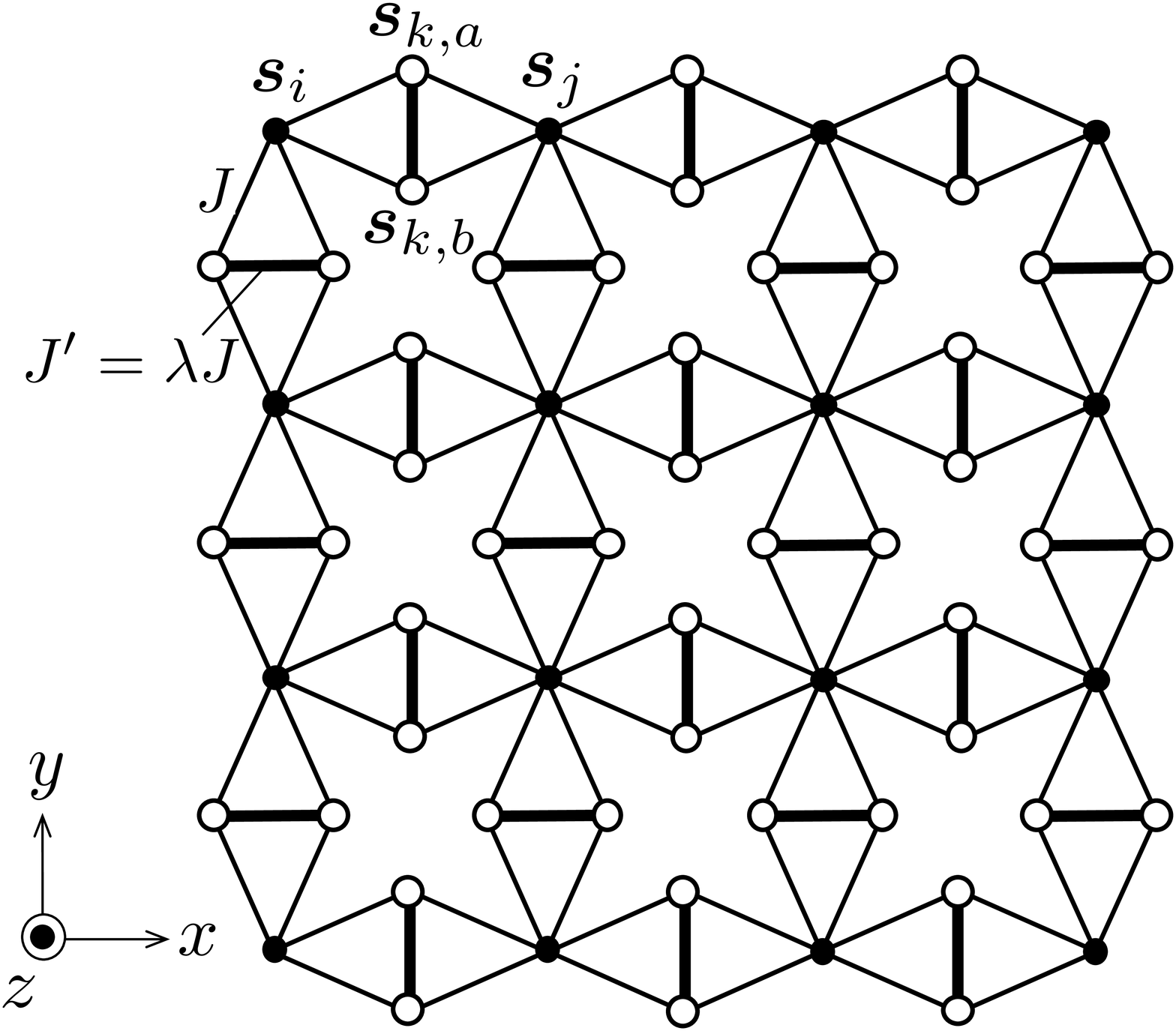}
\end{center}
\caption{Structure of the diamond-like-decorated square lattice. The thin and thick solid lines represent the antiferromagnetic interactions $J$ and $J'=\lambda J$, respectively. We call $\bm{s}_i$ and $\bm{s}_j$ the edge spins and the pair ($\bm{s}_{k,a}, \bm{s}_{k,b}$) a bond spin-pair. The edge spins and bond spin-pairs are indicated by the closed and open circles, respectively. The magnitude of all spin operators is $1/2$.  
 }
\label{fig:1}
\end{figure}
\begin{figure}[h] 
\begin{center}
\includegraphics[width=.99\linewidth]{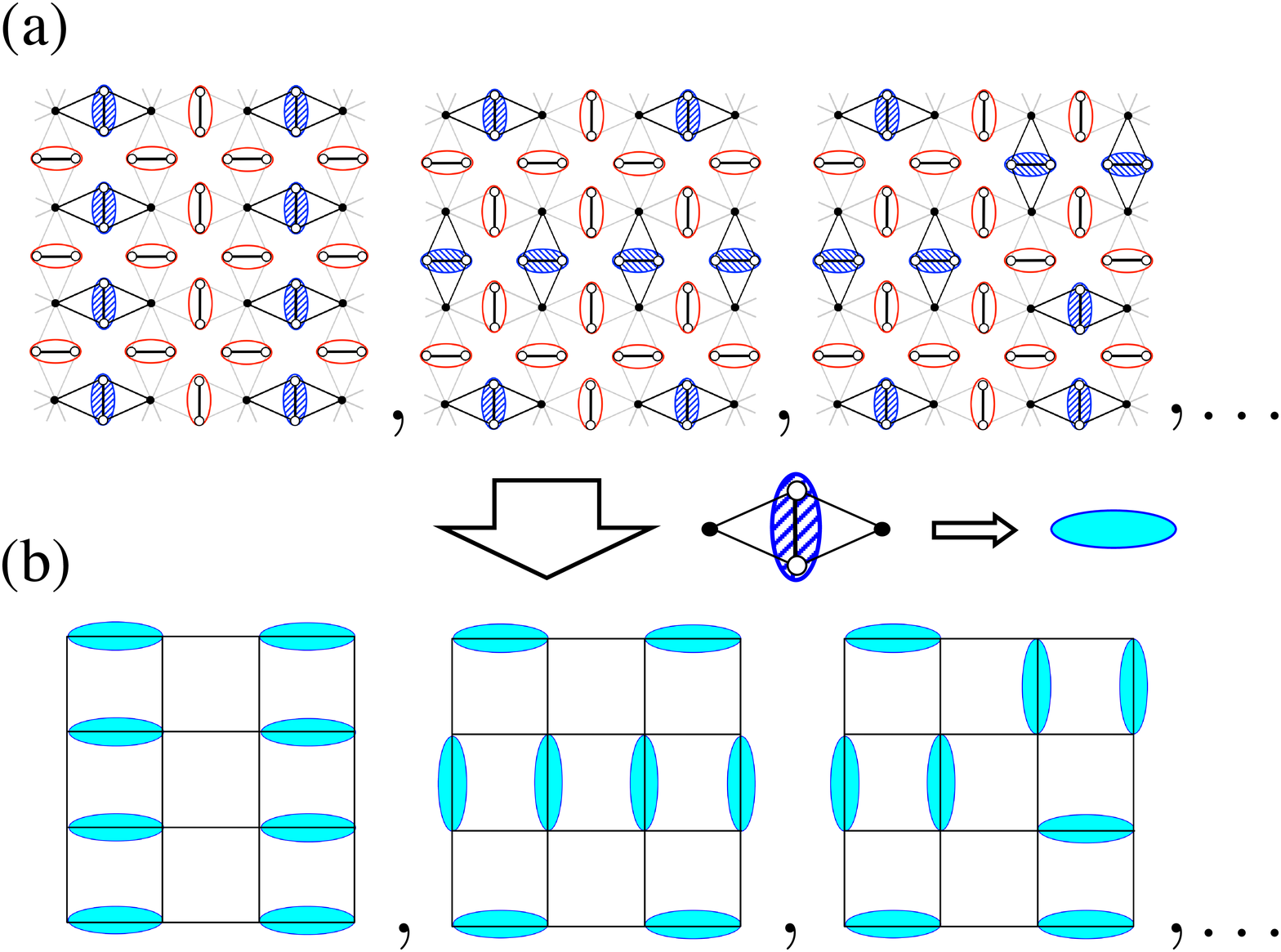}
\end{center}
\caption{(a) Macroscopically degenerated tetramer$-$dimer (MDTD) states for $0.974<\lambda<2$. The shaded blue and unshaded red ovals represent the triplet and singlet states on the bond spin-pair, respectively. (b) Square-lattice dimer-covering states when we regard a tetramer singlet as a ``dimer'' in the QDM.
 }
\label{fig:2}
\end{figure}

\section{Definition of the second-order effective Hamiltonian and the square-lattice QDM}
\label{sect:2}

The second-order effective Hamiltonian can be written as 
\begin{equation}
   H_{\rm{eff}}=-t \hat{T}+\epsilon_2\hat{D}_2+\epsilon_1\hat{D}_1+\epsilon_0\hat{D}_0,
\label{eq:4}   
\end{equation}
where $t$ represents the second-order pair-hopping amplitude and $\epsilon_2$, $\epsilon_1$, and $\epsilon_0$ represent the second-order perturbation energies when there are two, one, and zero dimers on a plaquette, respectively.
The operators $\hat{T}$, $\hat{D}_2$, $\hat{D}_1$, and $\hat{D}_0$ are defined by
\begin{align}
\hat{T}=&\sum \Bigl( \Bigr| \pa{} \Bigl\rangle \Bigl\langle\pb{}\Bigr|+\Bigr| \pb{} \Bigl\rangle \Bigl\langle\pa{}\Bigr| \Bigr),\notag \\
\hat{D}_2=&\sum \Bigl( \Bigr| \pa{} \Bigl\rangle \Bigl\langle\pa{}\Bigr|+\Bigr| \pb{} \Bigl\rangle \Bigl\langle\pb{}\Bigr| \Bigr),\notag \\
\hat{D}_1=&\sum \Bigl( \Bigr| \pc{} \Bigl\rangle \Bigl\langle\pc{}\Bigr|+\Bigr| \pd{} \Bigl\rangle \Bigl\langle\pd{}\Bigr| \\
&\;\;\;+\Bigr| \pe{} \Bigl\rangle \Bigl\langle\pe{}\Bigr|+\Bigr| \pf{} \Bigl\rangle \Bigl\langle\pf{}\Bigr|\Bigr),\notag\\
\hat{D}_0=&\sum  \Bigr| \pg{} \Bigl\rangle \Bigl\langle\pg{}\Bigr|.\notag
\label{eq:5}
\end{align}
Furthermore, Eqs. (\ref{eq:4}) can be rewritten as 
\begin{equation}
   H_{\rm{eff}}= -t \hat{T}+(\epsilon_2-2\epsilon_1+\epsilon_0)\hat{D}_2+\epsilon_1N
   \label{eq:9},
\end{equation}
where we use the condition $\hat{D}_2+\hat{D}_1+\hat{D}_0=N=\text{[total number of plaquettes]}$ and $\frac{1}{2}\left(2\hat{D}_2+\hat{D}_1\right)=\frac{N}{2}=\text{[total number of dimers]}$. 
The $\hat{D}_2$ coefficient on the right-hand side of Eq. (\ref{eq:9}) represents the dimer-dimer interaction
\begin{equation}
   v=\epsilon_2-2\epsilon_1+\epsilon_0,
\label{eq:10}   
\end{equation}
which represents repulsive ($v>0$) or attractive interaction ($v<0$) between dimers. 
Therefore, from Eqs. (\ref{eq:9}) and (\ref{eq:10}), we notice that the effective Hamiltonian $H_{\rm{eff}}$ can be written as the sum of $H_{\rm{QDM}}$ and the constant term $\epsilon_1N$,  which is the generation energy of a dimer.

\section{Calculation results for the dependence of $v/|t|$ on $\lambda$}
\label{sect:3}

This section presents calculation results for the dependence of $v/|t|$ on $\lambda$. 
First, we explain the method for introducing further neighbor couplings. 
As shown in Fig.~3, we introduce two kinds of further neighbor couplings $\Delta_{\rm{I}}$ and $\Delta_{\rm{II}}$, which are shown by the dashed lines and by the double dashes lines, respectively. 
The coupling $\Delta_{\rm{I}}$ connects between two adjacent diamond units, on the other hand, $\Delta_{\rm{II}}$ between two facing diamond units in the same plaquette, respectively. 

\begin{figure}[b] 
\begin{center}
\includegraphics[width=.75\linewidth]{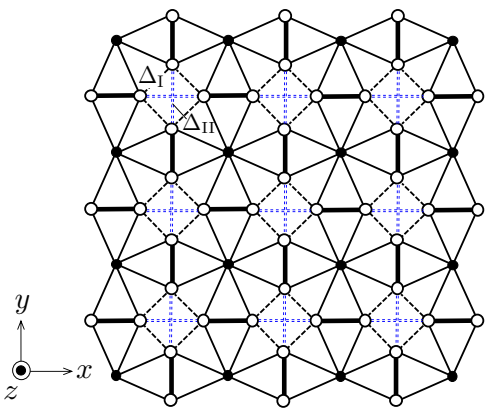}
\end{center}
\caption{Structure of the diamond-like-decorated square lattice with the introduction of two kinds of further neighbor couplings $\Delta_{\rm{I}}$ and $\Delta_{\rm{II}}$.
 }
\label{fig:3}
\end{figure}

\begin{figure}[b]
\begin{center}
\includegraphics[width=.99\linewidth]{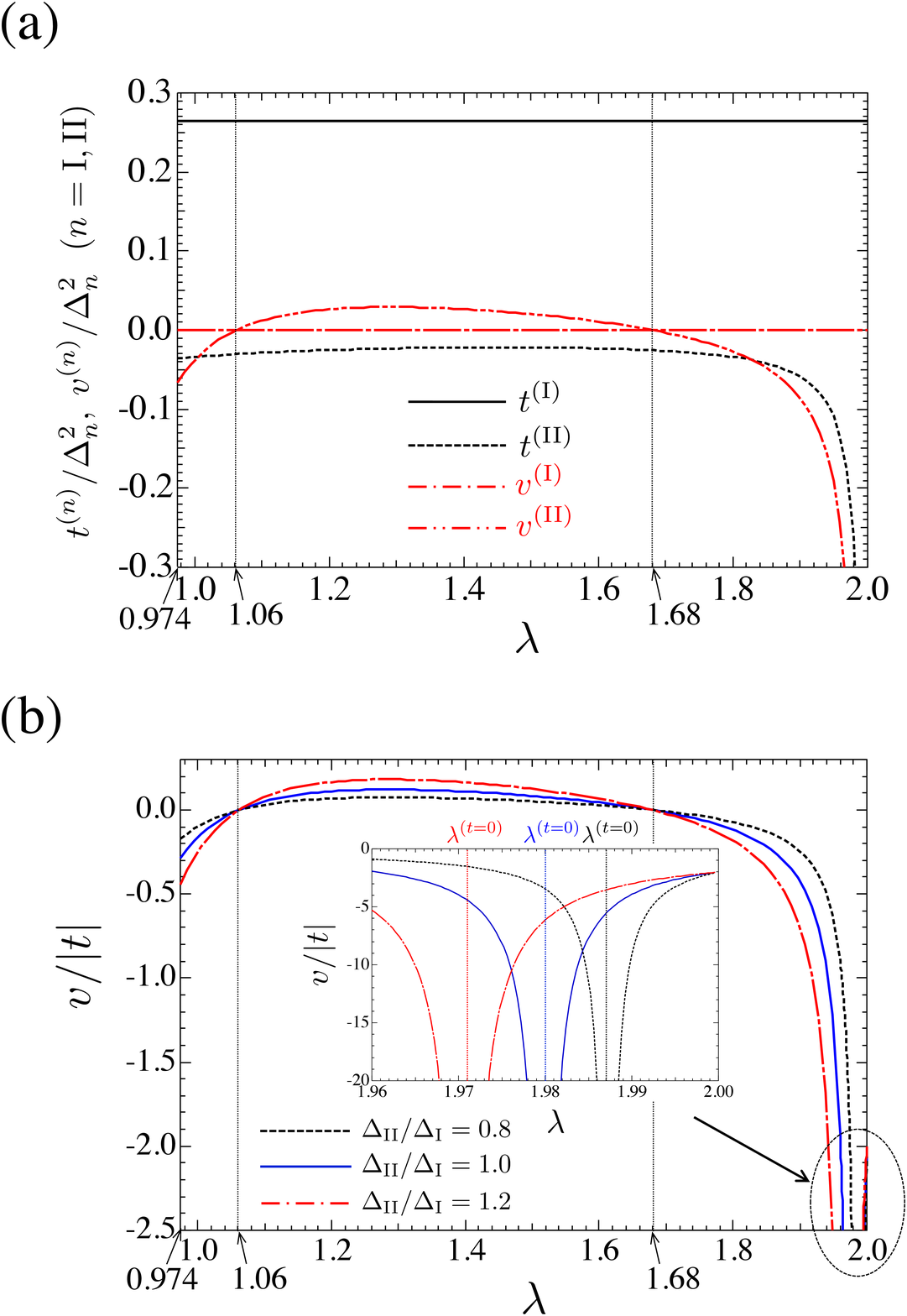}
\caption{Calculation results for dependence of (a) $v^{(\rm{I})}$, $t^{(\rm{I})}$, $v^{(\rm{II})}$, $t^{(\rm{II})}$, and (b) $v/|t|$ on $\lambda$. The inset shows an enlarged plot around $\lambda=\lambda^{(t=0)}$.
 }
\label{fig:4}
\end{center}
\end{figure}

In Fig.~\ref{fig:4}(a), we show numerical calculation results for the dependence of $v^{(\rm{I})}$, $t^{(\rm{I})}$, $v^{(\rm{II})}$, and $t^{(\rm{II})}$ on $\lambda$, where $v^{(\rm{I})}$ ($v^{(\rm{II})}$) and $t^{(\rm{I})}$ ($t^{(\rm{II})}$) represent the second-order perturbation matrix elements for the  dimer-dimer interaction and pair-hopping amplitude when we introduce the further neighbor couplings $\Delta_{\rm{I}}$ ($\Delta_{\rm{II}}$), respectively.
The horizontal axis shows $0.974<\lambda<2$, where the MDTD states are stabilized and the square lattice dimer-covering states are constructed. 
The dimer-dimer interaction $v^{(\rm{I})}$ becomes zero for all $\lambda$ in this region, which is based on the fact that the perturbation process contributions cancel each other out.\cite{Hirose1} 
On the other hand, the dimer-dimer interaction $v^{(\rm{II})}$ becomes $v^{(\rm{II})}\neq0$.  
We obtain $v^{(\rm{II})}>0$ ($v^{(\rm{II})}<0$), which is the repulsive (attractive) interaction between dimers, in the regions of $1.06<\lambda<1.68$ ($0.974<\lambda<1.06$ and $1.68<\lambda<2$). 
Note that, in the neighborhood of $\lambda=2$, we obtain large attractive interaction $v^{(\rm{II})}$, which originates from the fact that the energy gain of a plaquette with two dimers is larger than the others, i.e., $|\epsilon_2|\gg|\epsilon_0|, |\epsilon_1|$.
Furthermore, the $v^{(\rm{I})}=0$ and $v^{(\rm{II})}>0$ ($v^{(\rm{II})}<0$) results indicate that coupling $\Delta_{\rm{II}}$ produces repulsive (attractive) interaction instead of the coupling $\Delta_{\rm{I}}$. 
Next, focusing on the pair-hopping amplitude, we obtain $t^{(\rm{I})}=0.265\Delta_{\rm{I}}^2>0$, which is independent of $\lambda$. 
On the other hand, $t^{(\rm{II})}<0$ is obtained, which does depend on $\lambda$. 
The $\lambda$ dependence originates from the difference between the number of dimers in the intermediate state and that in the initial (final) state during the perturbation process.
If the numbers of dimers are the same in both the initial (final) and the intermediate states, there is no $\lambda$ dependence. 
On the other hand, if the numbers of dimers are different in these states, there is $\lambda$ dependence. 
We describe the details of these dependences in Ref~.5. 
Furthermore, both $v^{(\rm{II})}$ and $t^{(\rm{II})}$ diverge to $-\infty$ for $\lambda=2$, which is a phase transition point in the original spin Hamiltonian,\cite{Hirose3} and the energy denominator becomes zero. 

In Fig.~\ref{fig:4}(b), we show numerical calculations for the dependence of $v/|t|$ on $\lambda$ when $\Delta_{\rm{II}}/\Delta_{\rm{I}}=0.8, 1.0,$ and $1.2$, where $v=v^{(\rm{I})}+v^{(\rm{II})}(=v^{(\rm{II})})$ and $|t|=|t^{(\rm{I})}+t^{(\rm{II})}|$ are total second-order perturbation matrix elements when both couplings $\Delta_{\rm{I}}$ and $\Delta_{\rm{II}}$ are introduced. 
Note that the sign of the hopping parameter $t$ can be absorbed into an adequate phase factor for the dimer-covering states.  
From Fig.~\ref{fig:4}(b), we can see that the magnitude of $v/|t|$ increases as $\Delta_{\rm{II}}/\Delta_{\rm{I}}$ becomes large, except in the neighborhood of $\lambda=2$.  
Furthermore, we define $\lambda$ as $\lambda^{(t=0)}$ when $t^{(\rm{I})}=|t^{(\rm{II})}|$ $(t=0)$ is obtained.
Thus, when $\lambda<\lambda^{(t=0)}$ $(\lambda>\lambda^{(t=0)})$, we obtain $v/|t|\to-\infty$ for $\lambda\to\lambda^{(t=0)}-0$ ($\lambda\to\lambda^{(t=0)}+0$). 
When $0.974<\lambda<1.06$ and $1.68<\lambda<2$, we suggest that our obtained results correspond to the columnar phase of the square-lattice QDM because $v/|t|\le0$ is obtained, i.e., there is an attractive interaction between dimers. 
On the other hand, when $1.06<\lambda<1.68$, we obtain $v/|t|>0$ and in the case of $\Delta_{\rm{I}}=0$ and $\Delta_{\rm{II}}\neq0$, we obtain $v/|t|=v^{(\rm{II})}/|t^{(\rm{II})}|=1$ at $\lambda=1.18$ and $1.48$, which shows realization of the {\it{RVB state}}.\cite{Rokhsar1988}

Note that, in Fig.~3, we set the direction of bond spin-pairs parallel to the plane formed by the edge spins, i.e., we set the direction of bond spin-pairs orthogonal to the $z$-axis when we introduce the further neighbor couplings $\Delta_{\rm{I}}$ and $\Delta_{\rm{II}}$. 
However, we can consider another method of introducing the further neighbor couplings. 
In our previous studies\cite{Hirose2}, we set the direction of the bond spin-pairs orthogonal to the plane formed by the edge spins, i.e., we set the direction of the bond spin pairs parallel to the $z$-axis. As a result, we obtained $v^{(\rm{I})}=0$ and $v^{(\rm{II})}<0$ for all $\lambda$. 
Therefore, we suggested that the columnar phase is stabilized for all $\lambda$.


\section{Conclusions}
\label{sect:4}

We derived a square-lattice QDM as a second-order effective Hamiltonian for spin-1/2 Heisenberg antiferromagnets on a diamond-like-decorated square lattice by introducing further neighbor couplings $\Delta_{\rm{I}}$ and $\Delta_{\rm{II}}$. 
One of the most interesting results is that, at $\lambda=1.18$ and $1.48$, the RVB state can be realized.  

The present model contains only quadratic Heisenberg-type couplings and is based on an ordinary two-dimensional lattice periodicity and connectivity. Therefore, our construction of QDMs from Heisenberg models provides a clear path towards experimental realization or material design of QDMs. 
For example, layered compounds of $\rm{Cu^{2+}}$ ions bridged by bidentate di-radical ligands, or alkali metal atoms on an optical superlattice, are accessible.\cite{Bloch1,Bloch2} 
New insights into progress in RVB physics could be gained if these materials could be synthesized.


%
%

%

\begin{acknowledgments}
This work was supported by JSPS KAKENHI Grant Numbers JP17J05190 and JP17K05519.
\end{acknowledgments}

\bibliography{your-bib-file}

\begin{thebibliography}{99}
\bibitem{Anderson1973} P.~W.~Anderson, Mater. Res. Bull. \textbf{8}, 153 (1973).
\bibitem{Rokhsar1988} D. S. Rokhsar and S. A. Kivelson, Phys. Rev. Lett. \textbf{61}, 2376 (1988).
\bibitem{book} \textit{Introduction to Frustrated Magnetism}, ed. C. Lacroix, P. Mendels, and F. Mila, Springer Series in Solid-State Sciences, (Springer, Heidelberg, 2011) Vol. 164.
\bibitem{Hirose1} Y. Hirose, A. Oguchi, and Y. Fukumoto, J. Phys. Soc. Jpn. \textbf{85}, 094002 (2016).
\bibitem{Hirose2} Y. Hirose, A. Oguchi, and Y. Fukumoto, J. Phys. Soc. Jpn. \textbf{86}, 124002 (2017), Erratum: J. Phys. Soc. Jpn. \textbf{87}, 048001 (2018).
\bibitem{Strecka}L. $\check{\rm C}$anov$\acute{\rm a}$ and J. Stre$\check{\rm c}$ka, Phys. Status Solidi B \textbf{247}, 433 (2010).
\bibitem{Hirose3} Y. Hirose, A. Oguchi, and Y. Fukumoto, J. Phys. Soc. Jpn. \textbf{86}, 014002 (2017).
\bibitem{Morita} K. Morita and N. Shibata, J. Phys. Soc. Jpn. \textbf{85}, 033705 (2016).
\bibitem{Bloch1} I. Bloch, J. Dalibard, and W. Zwerger, Rev. Mod. Phys. \textbf{80}, 885 (2008).
\bibitem{Bloch2} S. Nascimb$\grave{\rm{e}}$ne, Y.-A. Chen, M. Atala, M. Aidelsburger, S. Trotzky, B. Paredes, and I. Bloch, Phys. Rev. Lett. \textbf{108}, 205301 (2012).
\end{thebibliography}

\end{document}